# Encryption Quality Analysis and Security Evaluation of CAST-128 Algorithm and its Modified Version using Digital Images

Krishnamurthy G N,

Dr. V Ramaswamy

*Abstract*— **this paper demonstrates analysis of well known block cipher CAST-128 and its modified version using avalanche criterion and other tests namely encryption quality, correlation coefficient, histogram analysis and key sensitivity tests.**

## I. INTRODUCTION

CAST-128[1], [2], [3] is a design procedure for symmetric encryption algorithm developed by Carlisle Adams and Stafford Tavares. CAST has a classical Feistel network (Fig. 1) consisting of 16 rounds and operating on 64-bit blocks of plaintext to produce 64-bit blocks of cipher text. The key size varies from 40 bits to 128 bits in 8-bit increments.

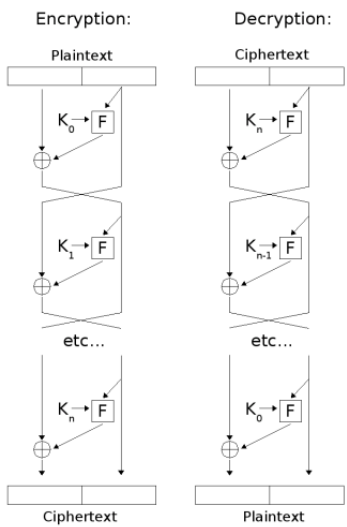

Fig. 1 CAST-128 Encryption and Decryption

### A. Function F

The function F (Fig. 2) is designed to have good confusion, diffusion and avalanche properties. It uses S-box substitutions, mod 2 addition and subtraction, exclusive OR operations and key dependent rotation. The strength of the F function is based primarily on the strength of the S-boxes.

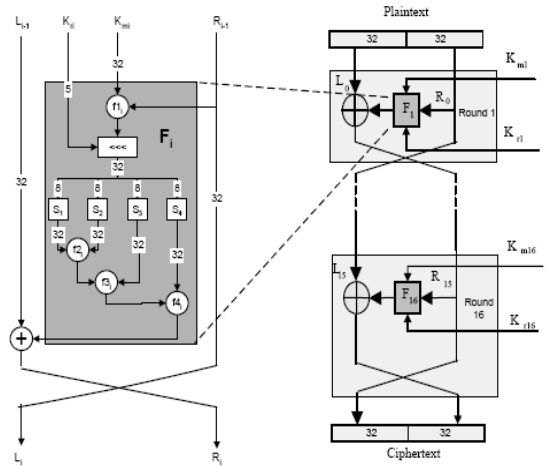

Further use of arithmetic, boolean and rotate operations add to its strength.

Fig. 2 CAST-128 Encryption Scheme

The function F includes the use of four S-boxes, each of size 8 x 32, the left circular rotation operation and four operation functions that vary depending on the round number. We label these operation functions as $f1_i$, $f2_i$, $f3_i$ and $f4_i$ (Fig. 2).

We use I to refer to the intermediate 32-bit value after the left circular rotation function and the labels $I_a$, $I_b$, $I_c$ and $I_d$ to refer to the 4 bytes of I where $I_a$ is the most significant and $I_d$ is the least significant. With these conventions, function F is defined as follows:

| Rounds | |
|---|---|
| 1,4,7,10,13,16 | $I = ((Km_i + R_{i-1}) <<< Kr_i)$ <br> $F = ((S_1[I_a]\ \hat{}\ S_2[I_b]) - (S_3[I_c])) + S_4[I_d]$ |
| 2,5,8,11,14 | $I = ((Km_i\ \hat{}\ R_{i-1}) <<< Kr_i)$ <br> $F = ((S_1[I_a] - S_2[I_b]) + (S_3[I_c]))\ \hat{}\ S_4[I_d]$ |
| 3,6,9,12,15 | $I = ((Km_i - R_{i-1}) <<< Kr_i)$ <br> $F = ((S_1[I_a] + S_2[I_b])\ \hat{}\ (S_3[I_c])) - S_4[I_d]$ |

*B. Modified Function*

Without violating the security requirements, CAST-128 function F can be modified[7] as follows:

| Rounds | $I = ((Km_i + R_{i-1}) <<< Kr_i)$ |
|---|---|
| 1,4,7,10,13,16 | $F = (S_1[I_a] \wedge S_2[I_b]) - (S_3[I_c] + S_4[I_d])$ |
| Rounds | $I = ((Km_i \wedge R_{i-1}) <<< Kr_i)$ |
| 2,5,8,11,14 | $F = (S_1[I_a] - S_2[I_b]) + (S_3[I_c] \wedge S_4[I_d])$ |
| Rounds | $I = ((Km_i - R_{i-1}) <<< Kr_i)$ |
| 3,6,9,12,15 | $F = (S_1[I_a] + S_2[I_b]) \wedge (S_3[I_c] - S_4[I_d])$ |

Note that the first and third operations of function F can be performed in parallel. For example, in round 1, $F = (S_1[I_a] \wedge S_2[I_b]) - (S_3[I_c] + S_4[I_d])$ in which operations $\wedge$ and $+$ can be done in parallel and the results of those two operations can be used to perform $-$ operation.

In [7] we have shown that this modification leads to 20% improvement in the exection time of function F.

Now we will show that the above modification to the function does not violate the security of the algorithm when compared to that of original algorithm. For this, we will make use of avalanche effect, encryption quality, key sensitivity test and statistical analysis.

## II. AVALANCHE EFFECT

We have used Avalanche effect[1], [2] to show that the modified algorithm also possesses good diffusion characteristics as that of original algorithm.

We have taken 60000 pairs of plaintexts with each pair differing only by one bit. We have encrypted them first by using the original algorithm and then by using modified one. For both the algorithms same key (K1) is used which is selected arbitrarily.

We have counted the number of times original algorithm gives better avalanche, the number of times the modified algorithm gives better avalanche and the number of times both algorithms give same avalanche. Tabulation of results observed by changing one bit of plaintext in the samples for rounds 2, 4, 6, 8, 10, 12, 14 and 16 of original and modified algorithms is as shown in table I.

Table I Avalanche Effect for one bit change in Plaintext

| Number of rounds | Number of pairs of plaintext samples | Number of times Original algorithm gives better Avalanche | Number of times Modified algorithm gives better Avalanche | Number of times both algorithms give same Avalanche |
|---|---|---|---|---|
| 2 | 60000 | 26413 | 25763 | 7824 |
| 4 | 60000 | 26301 | 25845 | 7854 |
| 6 | 60000 | 25775 | 26596 | 7629 |
| 8 | 60000 | 26303 | 25887 | 7810 |
| 10 | 60000 | 25790 | 26320 | 7890 |
| 12 | 60000 | 25748 | 26264 | 7988 |
| 14 | 60000 | 25622 | 26542 | 7836 |
| 16 | 60000 | 26017 | 26122 | 7861 |

Table II Avalanche Effect for one bit change in Key

| Number of rounds | Number of pairs of plaintext samples | Number of times Original algorithm gives better Avalanche | Number of times Modified algorithm gives better Avalanche | Number of times the both algorithms give same Avalanche |
|---|---|---|---|---|
| 2 | 60000 | 26186 | 25977 | 7837 |
| 4 | 60000 | 25773 | 26563 | 7664 |
| 6 | 60000 | 26249 | 25943 | 7808 |
| 8 | 60000 | 26405 | 25762 | 7833 |
| 10 | 60000 | 26067 | 26202 | 7731 |
| 12 | 60000 | 25782 | 26189 | 8029 |
| 14 | 60000 | 25950 | 26166 | 7884 |
| 16 | 60000 | 26204 | 25978 | 7818 |

We have carried out similar tests by changing one bit in the key and using set of 60000 plaintext samples. First we encrypted these plaintexts with a key using both the algorithms. Then just by changing the key by one bit chosen randomly the same set of plaintexts is encrypted using both the algorithms. We have observed the change in the number of bits. The results are tabulated as shown in table II for different rounds.

From the results, we can observe that both the algorithms posses good avalanche properties.

## III. ENCRYPTION QUALITY ANALYSIS

The quality of image encryption[6],[11] may be determined as follows:

Let $F$ and $F'$ denote the original image (plainimage) and the encrypted image (cipherimage) respectively each of size M*N pixels with L grey levels. $F(x, y), F'(x, y) \in \{0,..,L-1\}$ are the grey levels of the images $F$ and $F'$ at position $(x, y)$ ($0 \leq x \leq M-1, 0 \leq y \leq N-1$). Let $H_L(F)$ denote the number of occurrences of each grey level L in the original image (plainimage) $F$. Similarly, $H_L(F')$ denotes the number of occurrences of each grey level L in the encrypted image (cipherimage) $F'$. The encryption quality represents the average number of changes to each grey level L and is expressed mathematically as

$$\text{Encryption Quality} = \frac{\sum_{L=0}^{255} | H_L(F') - H_L(F) |}{256}$$

For all tests we have used two images Ape.bmp and Cart.bmp both of size 512x512.

The effect of number of rounds r on the encryption quality for CAST-128 and Modified CAST-128 is investigated. The encryption quality of CAST-128 and modified CAST-128 is computed as a function of number of rounds (r) using Ape.bmp as plainimage and its corresponding encrypted images. The results are tabulated for different rounds as shown in tables III. This procedure is repeated for another bitmap image Cart.bmp and the results are shown in table IV.

Table III  Encryption Qualities using Ape.bmp as Plainimage

| Encryption Quality (E.Q) of CAST-128 and Modified CAST-128 | | |
|---|---|---|
| Number of Rounds r | Algorithm type | |
| | CAST-128 | Modified CAST-128 |
| 2 | 766.242188 | 770.476562 |
| 4 | 765.648438 | 766.226562 |
| 6 | 766.281250 | 768.320312 |
| 8 | 765.859375 | 762.335938 |
| 10 | 766.093750 | 764.445312 |
| 12 | 763.812500 | 765.875000 |
| 14 | 763.718750 | 765.757812 |
| 16 | 766.375000 | 769.203125 |

Table IV  Encryption Qualities using Cart.bmp as Plainimage

| Encryption Quality (E.Q) of CAST-128 and Modified CAST-128 | | |
|---|---|---|
| Number of Rounds r | Algorithm type | |
| | CAST-128 | Modified CAST-128 |
| 2 | 1204.046875 | 1203.859375 |
| 4 | 1205.562500 | 1015.625000 |
| 6 | 1201.250000 | 1187.273438 |
| 8 | 1197.726562 | 1199.000000 |
| 10 | 1202.265625 | 1217.406250 |
| 12 | 1223.007812 | 1199.453125 |
| 14 | 1192.523438 | 1196.585938 |
| 16 | 1204.617188 | 1204.101562 |

The above results show that modification done to the function does not degrade the quality of encryption.

## IV. KEY SENSITIVITY TEST

We have conducted key sensitivity test[6], [11] on the image Cart.bmp for original and modified CAST-128 algorithms using the following 128 bit keys K1 and K2 where K2 is obtained by complementing one of the 128 bits of K1 which is selected randomly. The hexadecimal digits of K1 and K2 which have this difference bit are shown in bold case.
K1 = ADF2**7**8565E262AD1F5DEC94A0BF25B27 (Hex)
K2 = ADF2**3**8565E262AD1F5DEC94A0BF25B27 (Hex)

First the plainimage Cart.bmp (Fig. 3A) is encrypted with K1 using original CAST-128 algorithm and then by using K2. These cipher images are shown in Fig. 3B and 3C. Then we have counted the number of pixels that differ in the encrypted images. The result is **99.610687**% of pixels differ from the image encrypted with the key K2 from that encrypted with K1. The difference image shown in 3D confirms this. When we tried to decrypt the image which is encrypted with K1 using K2 (Fig. 3E), or vice-versa (Fig. 3F) no original information is revealed.

Above experiment is repeated for modified CAST-128. **99.602608**% of pixels differ from the image encrypted with K1(Fig. 4B) compared to the image encrypted with K2 (Fig. 4C). Fig. 4D shows the difference of the two images. When we tried to decrypt images encrypted with K1 and K2 by using keys K2 and K1 respectively decryption completely failed as it has happened in original CAST-128 and the results are shown in 4E and 4F.

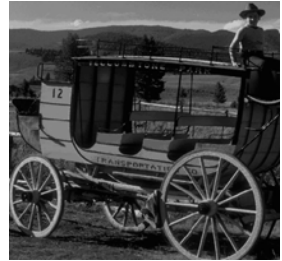
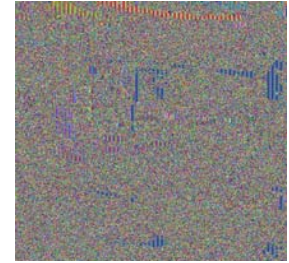

Fig. 3A Plainimage Cart.bmp
Fig. 3B Encrypted with Key K1

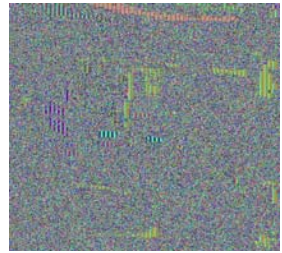
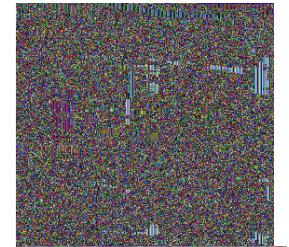

Fig. 3C Encrypted with Key K2
Fig. 3D Difference of 3B & 3C

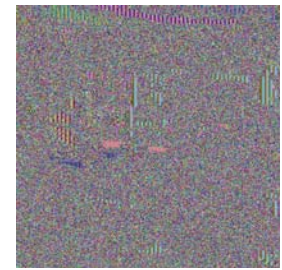

Fig. 3E Encrypted with Key K1 but Decrypted with K2   Fig. 3F Encrypted with Key K2 but Decrypted with K1

Fig. 3 Results of Key Sensitivity Test for Original CAST-128 Algorithm

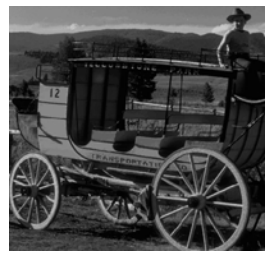
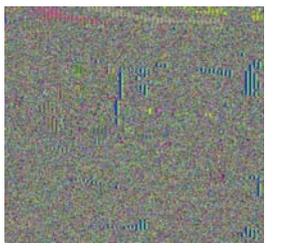

Fig. 4A Plainimage Cart.bmp
Fig. 4B Encrypted with Key K1

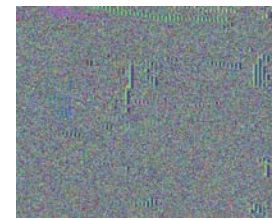
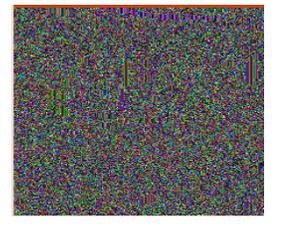

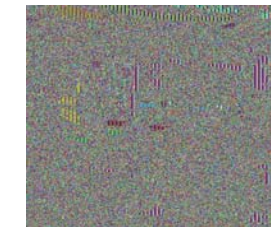
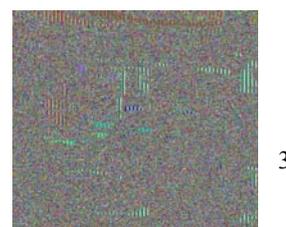

Fig. 4C Encrypted with Key K2   Fig. 4D Difference of 4B & 4C   Original Image (Ape.bmp)

Fig. 4E Encrypted with Key K1 but Decrypted withK2  Fig. 4F Encrypted with Key K2 but Decrypted with K1

Fig.4 Results of Key Sensitivity Test for Modified CAST-128 Algorithm

The textures visible in the cipherimages of the above tests is an indication of appearance of large area in the original image where pixel values rarely differ. It is the property of block ciphers that for a given input there will be fixed ciphertext, which means as long as plaintext block repeats, ciphertext block also repeats. This can be avoided by using one the modes of operation other than ECB mode.

## V. STATISTICAL ANALYSIS

This is shown by a test on the histograms[6], [11] of the enciphered images and on the correlations of adjacent pixels in the ciphered image.

### A. Histograms of Encrypted Images

We have selected Ape.bmp image as plainimage for histogram analysis. We have encrypted this image first by using original CAST-128 algorithm and then by using modified CAST-128 algorithm. Then we have generated histograms for plainimage and its encrypted images.

Fig. 5 shows the histograms for original image and its corresponding cipherimage obtained using original CAST-128 algorithm. Fig. 6 shows histogram for cipherimage encrypted using modified CAST-128 algorithm. From these figures we can see that the histogram of the encrypted images is fairly uniform and is significantly different from that of the original image.

From the histogram we can also observe that there is a huge variation in the percentage of number of pixels with a certain grey scale value which is varying from 0 to 1%. For cipher images this percentage is almost constant. This shows that the number of pixels with a certain grey scale value is almost same which is around 0.4% approximately. This is clearly shown in Fig. 5B, 5D and 6B.

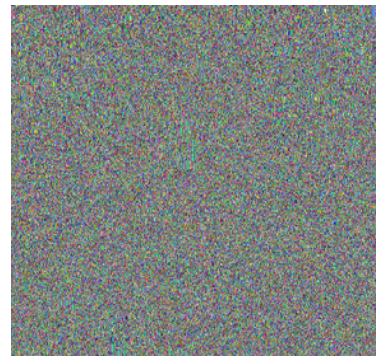

Fig. 5B Histogram for Original Image

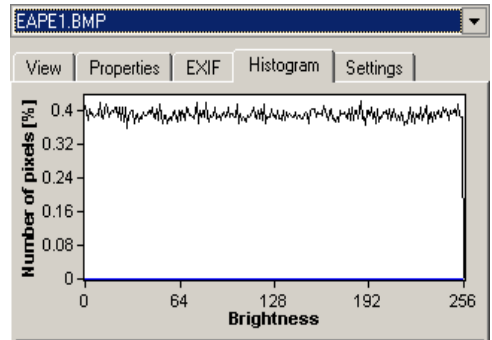

Fig. 5C Encrypted Image of Ape.bmp

Fig. 5D Histogram for Encrypted Image

Fig. 5 Histograms for Plainimage and Cipherimage of Original CAST-128 Algorithm

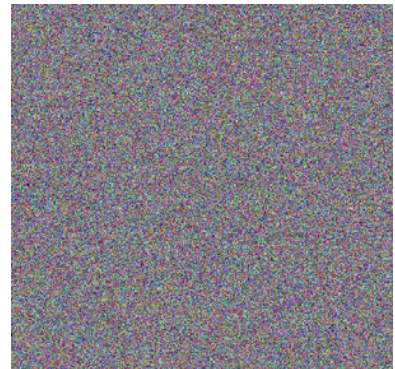

Fig. 6A Encrypted Image of Ape.bmp

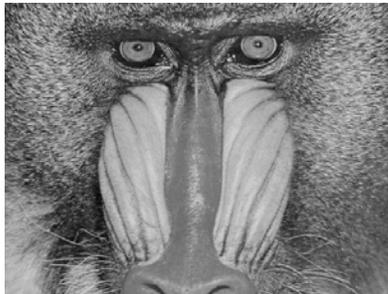

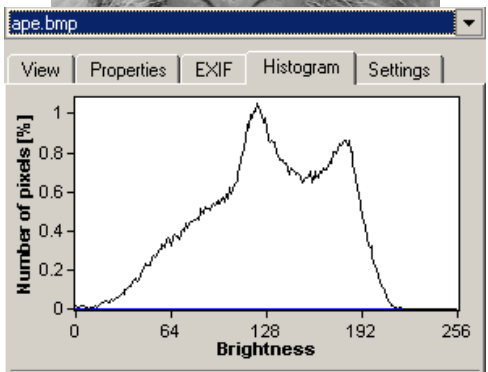

Fig. 5A

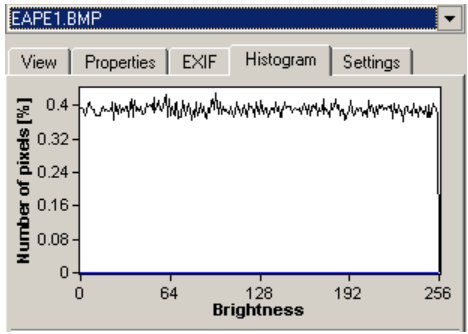

Fig. 6B Histogram for Encrypted Image

Fig. 6 Results of Histogram for Cipherimage of Modified CAST-128 Algorithm

*B. Correlation of Two Adjacent Pixels*

To determine the correlation between horizontally adjacent pixels[6], [11] in an image, the procedure is as follows:

First, randomly select N pairs of horizontally adjacent pixels from an image. Compute their correlation coefficient using the following formulae

$$E(x) = \frac{1}{N}\sum_{i=1}^{N} x_i$$

$$D(x) = \frac{1}{N}\sum_{i=1}^{N} (x_i - E(x))^2,$$

$$\text{cov}(x, y) = \frac{1}{N}\sum_{i=1}^{N} (x_i - E(x))(y_i - E(y)),$$

$$r_{xy} = \frac{\text{cov}(x, y)}{\sqrt{D(x)}\sqrt{D(y)}},$$

where x and y represent grey-scale values of horizontally adjacent pixels in the image. E(x) represents the mean of x values, D(x) represents the variance of x values, cov(x,y) represents covariance of x and y and $r_{xy}$ represents correlation coefficient.

We have randomly selected 1200 pairs of two adjacent pixels from the plainimage, Ape.bmp and the corresponding cipherimages encrypted using original and modified algorithms. Then we have computed the correlation coefficient using the above equations.

The correlation coefficient for plainimage was found to be **0.874144**. For cipherimage which is encrypted using original CAST-128, it is **0.016693** and it is **0.012245** for image encrypted using modified CAST-128. Fig. 7, 8 and 9 show the correlation distribution of two horizontally adjacent pixels for plainimage Ape.bmp and the encrypted images encrypted using original and modified CAST-128 algorithms respectively.

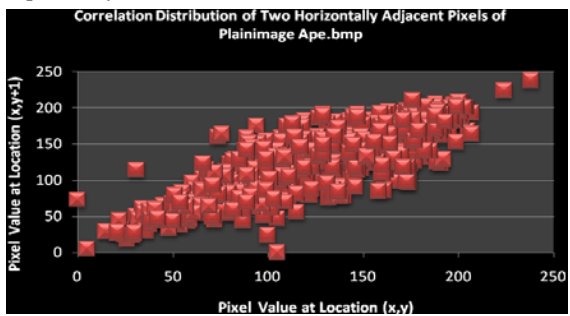

Fig. 7 Correlation Distribution of Two Horizontally Adjacent Pixels for Plainimage Ape.bmp

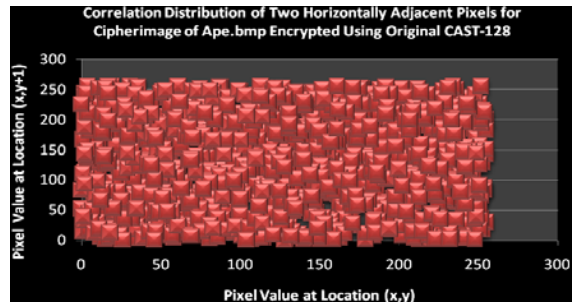

Fig. 8 Correlation Distribution of two Horizontally Adjacent Pixels for Cipherimage of Ape.bmp Encrypted using Original CAST-128 Algorithm

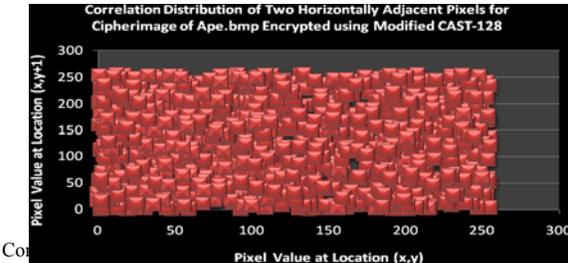

Fig. 9 Correlation Distribution of two Horizontally Adjacent Pixels for Cipherimage of Ape.bmp Encrypted using Modified CAST-128 Algorithm

Table V  Correlation Coefficients for Ape.bmp, Cart.bmp and their Encrypted Images

| Image Name | Plainimage | Correlation Coefficient for | | | |
|---|---|---|---|---|---|
| | | Original CAST-128 | | Modified CAST-128 | |
| | | Image Encrypted with key K1 | Image Encrypted with key K2 differing by 1 bit from K1 | Image Encrypted with key K1 | Image Encrypted with key K2 differing by 1 bit from K1 |
| Ape.bmp | 0.874144 | 0.016693 | 0.018121 | 0.012245 | 0.004573 |
| Cart.bmp | 0.919283 | 0.028416 | 0.045663 | 0.031307 | 0.019371 |

Table V gives the correlation coefficients for two bit map images Ape and Cart and their encrypted images using original and modified CAST-128 algorithms. The correlation coefficient values for plainimages are much larger than for that of encrypted images in both cases.

All the observations from the tests we conducted reveal a fact that the modified algorithm is at least as strong as original one.

## VI. CONCLUSION

We have made an attempt to analyse the security of original and modified versions of CAST-128 algorithm. We have also tried to demonstrate that the modification made to the function does not violate the security and is at least as strong as the original algorithm. For this purpose, we have used avalanche criterion, encryption quality, histogram analysis, key sensitivity test and correlation coefficient.

## REFERENCES

[1] B. Schneier, *"Applied Cryptography – Protocols, algorithms, and source code in C",* John Wiley & Sons, Inc., New York, second edition, 1996.

[2] William Stallings, "*Cryptography and Network Security"*, Third Edition, Pearson Education, 2003.

[3] *C.M. Adams, "Constructing symmetric ciphers using the CAST design procedure",* Designs, Codes, and Cryptography, Vol. 12, No. 3, November 1997, pp. 71–104.

[4] *Adams C, "The CAST-128 Encryption Algorithm*", RFC 2144, May 1997.

[5] Harley R. Myler and Arthur R. Weeks, *"The Pocket Handbook of Image Processing Algorithms in C",* Prentice-Hall, New Jersey, 1993.

[6] *Hossam El-din H. Ahmed, Hamdy M. Kalash, and Osama S. Farag Allah, "Encryption Quality Analysis of RC5 Block Cipher Algorithm for Digital Images",* Journal of Optical Engineering, vol. 45, 2006.

[7] *Kishnamurthy G.N, Dr. V Ramaswamy, "Performance Enhancement of CAST-128 algorithm by modifying its function"*, Proceedings of International Conference in CISSE 2007, University of Bridgeport, Bridgeport, CT, USA.

[8] *Krishnamurthy G N, Dr. V Ramaswamy "Encryption quality analysis and Security Evaluation of Blow-CAST-Fish using digital images",* Communicated to International Journal of Computational Science 2008.

[9] Osama S. Farag Allah, Abdul Hamid M. Ragib, and Nabil A. Ismali, *"Enhancements and Implementation of RC6 Block Cipher for Data Security",* IEEE Catalog Number: 01CH37239, Published 2001.

[10] Hossam El-din H. Ahmed, Hamdy M. Kalash, and Osama S. Farag Allah, *"An Efficient Chaos-Based Feedback Stream cipher (ECBFSC) for Image Encryption and Decryption",* Accepted for publication in An International Journal of Computing and Informatics, 2007.

[11] *Hossam El-din H. Ahmed, Hamdy M. Kalash. And Osama S. Farang Allah, "Encryption Effeciency Analysis and Security Evaluation of RC6 Block Cipher for Digital Images*",

International Journal Of Computer, Information , and System Science, and Engineering volume 1 number 1  2007 ISSN 1307-2331. pp 33-38.

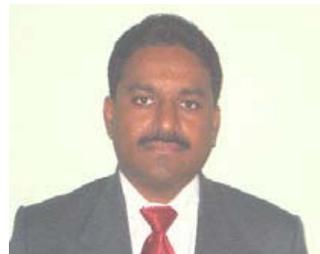

**Krishnamurthy G N** born on 15th March 1974, at Davangere, India. He has applied for membershipof IAENG. He obtained his B.E. degree in Electronics & Communication Engineering from Kuvempu University in 1996 and M.Tech. degree in Computer Science & Engineering from Visveswaraya technological University, India in 2000. He is presently pursuing his Ph.D. from Visveswaraya Technological University, India under the guidance of Dr. V Ramaswamy.

He has published papers in national and international conferences, journals in the area of Cryptography. After working as a lecturer (from 1997) he has been promoted to Assistant Professor (from 2005), in the Department of Information Science & Engineering, Bapuji Institute of Engineering & Technology, Davangere, affiliated to Visveswaraya Technological University, Belgaum, India. His area of interest includes Design and analysis of Block ciphers.

He is a life member of Indian Society for Technical Education, India.

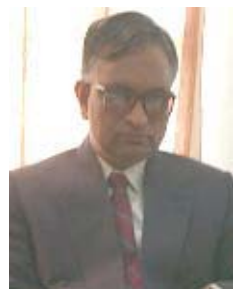

**Dr. V Ramaswamy** obtained his Ph.D. degree from Madras University, in 1982. He has applied for the membership of IAENG.

He is working as Professor and Head in the Department of Information Science and Engineering. He has more the 25 years of teaching experience including his four years of service in Malaysia. He is guiding many research scholars and has published many papers in national and international conference and in many international journals and authored one book. He has visited many universities in USA and Malaysia.

He is a life member of Indian Society for Technical Education, India.